\documentclass{ws-procs961x669}            % default, citations in superscript

\graphicspath{ {./images/} }

\defcitealias{soldateschi_2020}{SBD20}
\defcitealias{soldateschi_2021}{SBD21}

\begin{document}
\title{Quasi-universality of the magnetic deformation of neutron stars in general relativity and beyond}

\author{J. Soldateschi$^*$, N. Bucciantini and L. Del Zanna}

\address{Dipartimento di Fisica e Astronomia, Università degli Studi di Firenze, and INFN - Sezione di Firenze, Via G. Sansone 1,\\
I-50019 Sesto F. no (Firenze), Italy\\
*E-mail: jacopo.soldateschi@unifi.it}

\address{INAF - Osservatorio Astrofisico di Arcetri, Largo E. Fermi 5,\\
I-50125 Firenze, Italy}

\begin{abstract}
Neutron stars harbour extremely powerful magnetic fields, leading to their shape being deformed. Their magnetic deformation depends both on the geometry - and strength - of their internal magnetic field and on their composition, encoded by the equation of state. However, both the details of the internal magnetic structure and the equation of state of the innermost part of neutron stars are mostly unkown. We performed a study of numerical models of magnetised, static, axisymmetric neutron stars in general relativity and in one of its most promising extensions, scalar-tensor theories. We did so by using several realistic equations of state currently allowed by observational and nuclear physics constraints, considering also those for strange quark stars. We show that it is possible to find simple relations among the magnetic deformation of a neutron star, its Komar mass, and its circumferential radius in the case of purely poloidal and purely toroidal magnetic configurations satisfying the equilibrium criterion in the Bernoulli formalism. These relations are quasi-universal, in the sense that they mostly do not depend on the equation of state. Our results, being formulated in terms of potentially observable quantities, could help to understand the magnetic properties of neutron stars interiors and the detectability of continuous gravitational waves by isolated neutron stars, independently of their equation of state. In the case of scalar-tensor theories, these relations depend also on the scalar charge of the neutron stars, thus potentially providing a new way to set constraints on the theory of gravitation.
\end{abstract}


\bodymatter

\section{Introduction}\label{sec:1}
The most compact material objects in the known Universe are neutron stars (NSs), which are also known to harbour extremely powerful magnetic fields, especially in the sub-class known as magnetars \citep{duncan_1992,thompson_1993,thompson_1995,thompson_1996}: the surface magnetic field of NSs has been found to be in the range $10^{8-12}$G for radio and $\gamma$-ray pulsars \citep{asseo_2002,spruit_source_2009,ferrario_magnetic_2015}, while estimates in magnetars reach $10^{15}$G \citep{olausen_2014,popov_origins_2016}.
%While magnetars represent only a small subset [just over 30 sources \citep{olausen_2014}] of the NS population compared to regular pulsars [a few thousand sources \citep{atnf_2005}], it is believed that they may actually exist as a significant fraction of the young NS population \citep{kaspi_2017}.
While the surface and magnetospheric magnetic fields of NSs can be probed through many different methods \citep{rea_2010,staubert_2019}, their internal magnetic field remains mostly unknown, both in its strength and geometry. Predictions expect that values up to $10^{16}$G and $10^{17-18}$G could be reached inside magnetars and newly-born proto-NSs, respectively \citep{del_zanna_chiral_2018,ciolfi_2019,franceschetti_2020}. On the other hand, while it is known that neither purely poloidal nor purely toroidal magnetic configurations are stable \citep{prendergast_equilibrium_1956,frieben_equilibrium_2012} and mixed configurations are more favoured \citep{ciolfi_2013,uryu_equilibrium_2014,pili_axisymmetric_2014}, the extact magnetic field geometry in the interior of NSs is less clear. Fortunately, these magnetic fields present potentially observable effects to the structure of NSs: they can modify the torsional oscillations of NSs \citep{samuelsson_2007,sotani_2015}, their cooling properties \citep{page_2004,aguilera_2008} and their deformation \citep{haskell_2008,gomes_2019}.

In this scenario, another unknown regarding the internal structure of NSs enters the game: the equation of state (EoS) of NSs remains mostly unconstrained. Recent observational results rejected the validity of many EoS: the observation of very massive NSs [e.g. the most massive NS known to date, with a mass potentially reaching $\sim 2.28$M$_\odot$ \citep{kandel_2020}]; the limits on the stiffness of NSs \citep{abbott_2018_1,Bauswein_2019} obtained by the first observation of gravitational waves (GWs) emitted by a binary NS merger \citep{abbott_gw170817:_2017}; the results of the NICER telescope \citep{pang_2021,zhang_2021} on the possible NSs radii. However, this uncertainity is further enhanced by the effect that the strong magnetic field buried inside NSs exherts on their composition, for example determining the presence of exotic particles or the existence of a superconducting phase \citep{ruderman_1995,zdunik_2013,costa_2014,cai_2015,Drago_Lavagno_Pagliara_Pigato_2016}. For these reasons, undertanding and constraining the interplay between the magnetic field of NSs and their EoS is fundamental in order to improve our knowledge of these objects.

Other than observations in the electromagnetic domain, GWs can help to probe the inner structure of NSs. Of particular importance are continuous GWs (CGWs), which are emitted, for example, by a time-varying deformation. Such scenario can happen e.g. in the case of magnetically-deformed NSs whose magnetic axis is not aligned with their rotation axis \citep{bocquet_rotating_1995,cutler_2002,gomes_2019}. Since magnetic fields found in regular pulsars are too low to cause a significant deviation from spherical symmetry \citep{haskell_2008}, newly-born proto-NSs and millisecond magnetars are the most promising sources of CGWs \citep{dallosso_2021}, along with millisecond pulsars which possess a superconducting core \citep{cutler_2002}.

While the problem of the inner composition and magnetic field of NSs is an interesting problem on its own, it is deeply intertwined with the long-standing quest for the definitive theory of gravity. Even if general relativity (GR) remains the best theory to model the gravitational interaction, it has long been known that our understanding of gravity through GR presents some issues on the galactic and cosmological scales \citep{papantonopoulos_modifications_2015}. While one possibility is to introduce a dark sector \citep{trimble_existence_1987,peebles_cosmological_2003}, another path is to introduce modification to GR, leading to alternative theories of gravity \citep{capozziello_extended_2011}. Among the possible alternatives to GR, scalar-tensor theories (STTs) - in which gravity is mediated also by a scalar field non-minimally coupled to the spacetime metric - have gathered a strong interest because of their simplicity, because they derive from some suggested theories of quantum gravity \citep{damour_runaway_2002}, in general they satisfy the weak equivalence principle, and they are free of the issues affecting other alternatives to GR \citep{defelice_2006,defelice_2010,bertolami_2016}. Of great importance in the development of STTs was the discovery of a non-perturbative strong field effect, exhibited in some of these theories, named `spontaneous scalarisation' \citep{damour_nonperturbative_1993}, which leads to strong deviations from GR in the vicinity of compact material objects (i.e. NSs), while it remains unconstrained in the weak-field limit. The presence of a scalar field leads to an enrichment of the physics of NSs, for example by causing the emission of additional modes of GWs \citep{eardley_1973,pang_2020} or by modifing the mass-radius relation of NSs, the binary NS merger dynamics \citep{shibata_coalescence_2014}, the frequency of NS normal modes \citep{sotani_2005}, their tidal and rotational deformation \citep{pani_2014,doneva_rapidly_2014}, their magnetic deformation \citep{soldateschi_2020,soldateschi_2021,soldateschi_2021_1} and the light propagation properties in their vicinity \citep{bucciantini_2020}. While massless STTs have been mostly ruled out by observations \citep{will_confrontation_2014,shao_constraining_2017,voisin_2020}, STTs with a massive scalar field are still viable \citep{ramazanoglu_spontaneous_2016,yaza_2016,rosca_2020}. As anticipated, part of the phenomenology of STTs is degenerate with the EoS of NSs, for example concerning the mass-radius relation or their deformabilities. For such reason, disentangling their roles is important in order to have a more informative interpretation of observations.
\\\\
In the following we assume a signature $\{-, +, +, +\}$ for the
spacetime metric and use Greek letters $\mu$, $\nu$, $\lambda$, ... (running from
0 to 3) for 4D spacetime tensor components, while Latin letters
$i$, $j$, $k$, ... (running from 1 to 3) are employed for 3D spatial tensor components. Moreover, we use the dimensionless units where $c = G = \mathrm{M}_\odot = 1$, and we absorb the $\sqrt{4\pi}$ factors in the definition of the electromagnetic quantities. All quantities calculated in the Einstein frame (E-frame) are denoted with a bar ($\bar{\cdot}$) while all quantities calculated in the Jordan frame (J-frame) are denoted with a tilde ($\tilde{\cdot}$).

\section{Neutron stars in general relativity}\label{sec:2}

The spacetime metric in the case of static, axisymmetric configurations can be well approximated \citep{oron_relativistic_2002,pili_general_2017} using the conformally flat condition (CFC) \citep{wilson_mathews_2003,isenberg_waveless_2008}. Then, for spherical-like coordinates $x^\mu = [t,r,\theta,\phi]$, the line element is
\begin{equation}\label{eq:3+1metric}
    g_{\mu \nu}\mathrm{d}x^\mu \mathrm{d}x^\nu = -\alpha^2\mathrm{d}t^2 + \psi^4\left[ \mathrm{d}r^2 + r^2\mathrm{d}\theta^2 + r^2\sin^2\theta \mathrm{d}\phi^2 \right] \; ,
\end{equation}
where $g_{\mu \nu}$ is the spacetime metric, with determinant $g$, $\alpha (r,\theta)$ is the lapse function and $\psi (r,\theta)$ is the conformal factor. The energy-momentum tensor for a magnetised ideal fluid is \citep{bucciantini_fully_2013,tomei2020}
\begin{equation}\label{eq:tmunu}
    T^{\mu \nu} = \left( \rho + \varepsilon + p \right) u^\mu u^\nu + p g^{\mu \nu} + F^\mu _\lambda F^{\nu \lambda} - \frac{1}{4}F^{\lambda \kappa} F_{\lambda \kappa} g^{\mu \nu} \; ,
\end{equation}
where $\rho$ is the rest mass density, $\varepsilon$ is the internal energy density, $p$ is the pressure, $u^\mu$ is the four-velocity and $F^{\mu \nu}$ is the Faraday tensor. The 3+1 formalism \citep{alcubierre_introduction_2008,gourgoulhon_3+1_2012} can be used to recast the equations in a more computationally-friendly way; under these assumptions, the Einstein equations for the metric become two Poisson-like equations, one for $\psi$ and one for $\alpha \psi$ \citep{pili_axisymmetric_2014}.

As for the magnetohydrodynamics (MHD) quantities, using the pseudo-enthalpy $h$, related to the pressure and density by ${\rm d} \ln{h} = {\rm d} p /(\rho+\varepsilon+p)$, Euler's equation becomes the `generalised Bernoulli integral' \citep{pili_axisymmetric_2014},
\begin{equation}\label{eq:bernoulli}
	\ln \left( \frac{h}{h_{\mathrm c}} \right) + \ln \left( \frac{ \alpha}{ \alpha _{\mathrm c}} \right) - \mathcal{M}=0 \; ,
\end{equation}
where $\mathcal{M}$ is the magnetisation function and $h_{\mathrm c}$ and $\alpha _{\mathrm c}$ are the values of $h$ and $\alpha$ at the centre of the star, respectively (having assumed $\mathcal{M}_{\mathrm c}=0$). The magnetisation function in the case of a purely poloidal magnetic field is assumed to be $\mathcal{M}=k_{\mathrm{pol}}A_\phi$, where $k_{\mathrm{pol}}$ is the poloidal magnetisation constant and $A_\phi$ is the $\phi$-component of the vector potential, computed by solving the relativistic Grad-Shafranov equation. For a purely toroidal magnetic field, we instead use $\mathcal{M} = - k_\mathrm{tor}^2 (\rho +\varepsilon + p) \mathcal{R}^2$,
where $k_{\mathrm{tor}}$ is the toroidal magnetisation constant and $\mathcal{R}^2~=~\alpha ^2 \psi ^4 r^2 \sin ^2 \theta$. In the first case, the poloidal components of the magnetic field are found through the Grad-Shafranov equation \citep{del_zanna_exact_1996,pili_general_2017} for $A_\phi$; in the second case, the toroidal component is proportional to $(\rho +\varepsilon + p) \mathcal{R}^2/\alpha$. Finally, the EoS closes the system of equations. We describe the EoS we used in Sect.~\ref{sec:6}. For a more detailed description of the equilibrium formalism the reader is referred to Refs.~\citenum{soldateschi_2020,soldateschi_2021,soldateschi_2021_1}.

\section{Scalar-tensor theories in a nutshell}\label{sec:3}

The action of massless STTs in the J-frame, according to the `Bergmann-Wagoner formulation' \citep{bergmann_comments_1968,wagoner_scalar-tensor_1970,santiago_2000}, is
\begin{equation}
\label{eq:joract}
	\tilde{S}= \frac{1}{16\pi}\int \mathrm{d}^4x \sqrt{-\tilde{g}}\left[ \varphi \tilde{R} - \frac{\omega (\varphi)}{\varphi} \tilde{\nabla} _\mu \varphi \tilde{\nabla} ^\mu \varphi \right]
	+ \tilde{S}_\mathrm{p}\left[ \tilde{\Psi} , \tilde{g}_{\mu \nu} \right]  \; ,
\end{equation}
where $\tilde{g}$ is the determinant of the spacetime metric $\tilde{g}_{\mu \nu}$, $\tilde{\nabla} _\mu$ its associated covariant derivative, $\tilde{R}$ its Ricci scalar, $\omega (\varphi)$ is the coupling function of the scalar field $\varphi$, and $\tilde{S}_\mathrm{p}$ is the action of the physical fields $\tilde{\Psi}$. In the E-frame, the action is obtained by making the conformal transformation $\bar{g}_{\mu \nu}= \mathcal{A}^{-2}(\chi) \tilde{g}_{\mu \nu}$, where $\mathcal{A}^{-2}(\chi) =\varphi (\chi)$ and $\chi$ is a redefinition of the scalar field in the E-frame, related to $\varphi$ by ${\mathrm d}\chi / {\mathrm d}\ln \varphi~=~\{[\omega{(\varphi)} +3]/4\}^{1/2}$. In the E-frame, the scalar field is minimally coupled to the metric: Einstein's field equations retain their usual form in the E-frame, but the energy-momentum tensor is now the sum of the one describing the fluid and electromagnetic fields and of the scalar field one. On the other hand, the scalar field is minimally coupled to the physical fields in the J-frame: the MHD equations in the J-frame have the same expression as in GR. In addition to the metric and MHD equations, in STTs we have an additional equation to solve for the scalar field. In the E-frame it is
\begin{equation}\label{eq:scal1}
	\Delta \chi = -4\pi \bar{\psi}^4 \alpha _\mathrm{s}(\chi) \mathcal{A}^4 \tilde{T}-\partial \ln \left( \bar{\alpha} \bar{\psi} ^2 \right) \partial \chi \; ,
\end{equation}
where $\Delta = f^{ij} \hat{\nabla} _i \hat{\nabla} _j$ and $\hat{\nabla} _i$ are, respectively, the 3D Laplacian and nabla operator of the flat space metric $f_{ij}$, $\partial f \partial g~=~\partial _r f \partial _r g + (\partial _\theta f \partial _\theta g)/r^2$, $\alpha _\mathrm{s}(\chi) = {\mathrm d} \ln \mathcal{A}/{\mathrm d}\chi$ and $\tilde{T} = 3\tilde{p}- \tilde{\varepsilon}-\tilde{\rho}$ is the trace of the J-frame energy momentum tensor of the fluid and electromagnetic fields. We used an exponential coupling function $\mathcal{A}(\chi)~=~\exp \{ \alpha _0 \chi + \beta _0 \chi ^2 /2 \} $ \citep{damour_nonperturbative_1993}. The $\alpha _0$ parameter regulates the effects of the scalar field in the weak-field limit, while the $\beta _0$ parameter controls spontaneous scalarisation. We chose $\alpha _0=-2\times 10^{-4}$ and $\beta _0\in [-6,-4.5]$. The most recent observational constraints to date require that, for massless scalar fields, $|\alpha _0| \lesssim 1.3\times 10^{-3}$ and $|\beta _0| \gtrsim 4.3$ \citep{voisin_2020}; in the case of massive scalar fields, lower values of $\beta _0$ are allowed \citep{doneva_rapidly_2016}, as long as the screening radius is smaller than the separation of the binary NSs whose observation confirmed the absence of dipolar GWs \citep{zhang_gravitational_2017,zhang_2019}. We emphasise that results found in a massless STT regarding the internal structure of a NS are also valid for STTs containing a screening effect as long as the screening radius is larger than the NS radius; as such, our results regarding the magnetic deformation of NSs are valid also in the case of a massive scalar field, with a mass such that its screening radius is larger than the NS radius but lower than the binary separation. For a more detailed description of the formulation of STTs within the 3+1 formalism the reader is referred to Refs.~\citenum{soldateschi_2020,soldateschi_2021,soldateschi_2021_1}.

\section{The magnetic deformation of polytropic neutron stars}\label{sec:4}

We now focus on the case of static NSs in the weak magnetic field regime, where the magnetic deformation of the star is well approximated by a perturbative approach; this was shown to happen for ${B}_\mathrm{max}\lesssim 10^{17}$G \citep{pili_general_2015,bucciantini_role_2015}, where $ B_\mathrm{max}=\max [\sqrt{ B_i  B^i}]$ and $ B_i$ are the components of the magnetic field. We focus only on the range of masses corresponding to stable configurations. All the results shown are computed using the \texttt{XNS} code \citep{bucciantini_general_2011,pili_axisymmetric_2014,pili_general_2015,pili_general_2017,soldateschi_2020,soldateschi_2021,soldateschi_2021_1}, which solves the coupled equations for the metric, scalar field, and MHD structure of a NS under the assumptions of stationarity and  axisymmetry, adopting conformal flatness and maximal slicing.

In Newtonian gravity \citep{wentzel_1960,ostriker_1969} and in GR \citep{frieben_equilibrium_2012,pili_general_2017}, in the limit of weak magnetic fields, the quadrupole deformation $e$ of a magnetised NS can be expressed as a linear function of $B^2_\mathrm{max}$ \citep{pili_general_2017}. Equivalently, instead of using $B^2_\mathrm{max}$ one can parametrise $e$ also in terms of $\mathcal{H}/W$, where $\mathcal{H}$ is the magnetic energy of the NS, defined in the J-frame as
\begin{equation}
\tilde{\mathcal{H}} = \pi \int \mathcal{A}^3\tilde{B}_i \tilde{B}^i \sqrt{\bar{\gamma}}\mathrm{d}r \mathrm{d}\theta \; ,
\end{equation}
$W$ being its binding energy.

In STTs, we found that $\bar{e}$ still has a linear trend with $\tilde{B}^2_\mathrm{max}$ (or $\tilde{\mathcal{H}}/\bar{W}$), but with coefficients that depend more strongly on the baryonic mass $M_0$ (which is the same in the J-frame and in the E-frame) than in GR; the strength of this dependence is related to the value of $\beta _0$. In the limit $\tilde{B}_\mathrm{max} \rightarrow 0$, keeping fixed $M_0$ and $\beta _0$:
\begin{equation}\label{eq:coeffs}
	|\bar{e}| = c_\mathrm{B} \tilde{B}^2_\mathrm{max} + \mathcal{O}\left( \tilde{B}^4_\mathrm{max} \right) , \quad |\bar{e}| = c_\mathrm{H} \frac{\tilde{\mathcal H}}{\bar{ W}} + \mathcal{O}\left( \frac{\tilde{\mathcal H} ^2}{\bar{ W} ^2} \right),
\end{equation}
where $c_\mathrm{B}=c_\mathrm{B}(M_0,\beta _0)$ and $c_\mathrm{H}=c_\mathrm{H}(M_0,\beta _0)$ are the `distortion coefficients', and  $\tilde{B}_\mathrm{max}$ is normalised to $10^{18}$G.

\begin{figure*}
    \centering
    \includegraphics[width=0.491\textwidth]{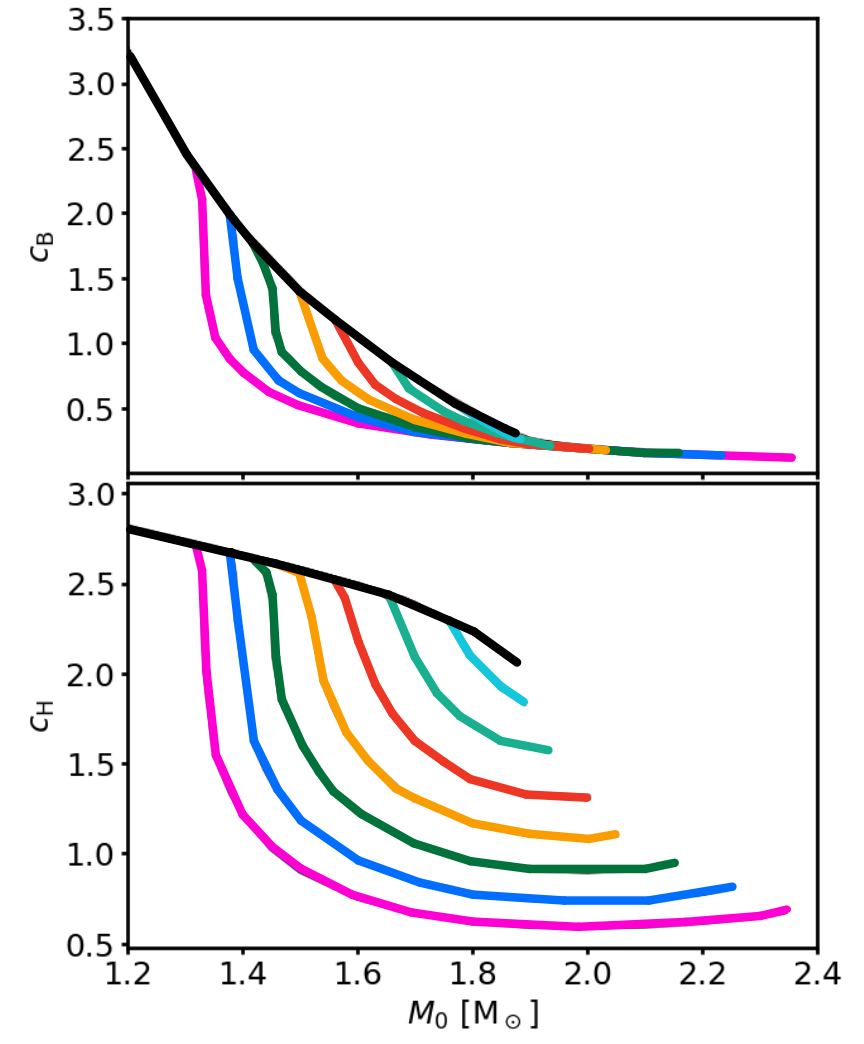}
	\includegraphics[width=0.489\textwidth]{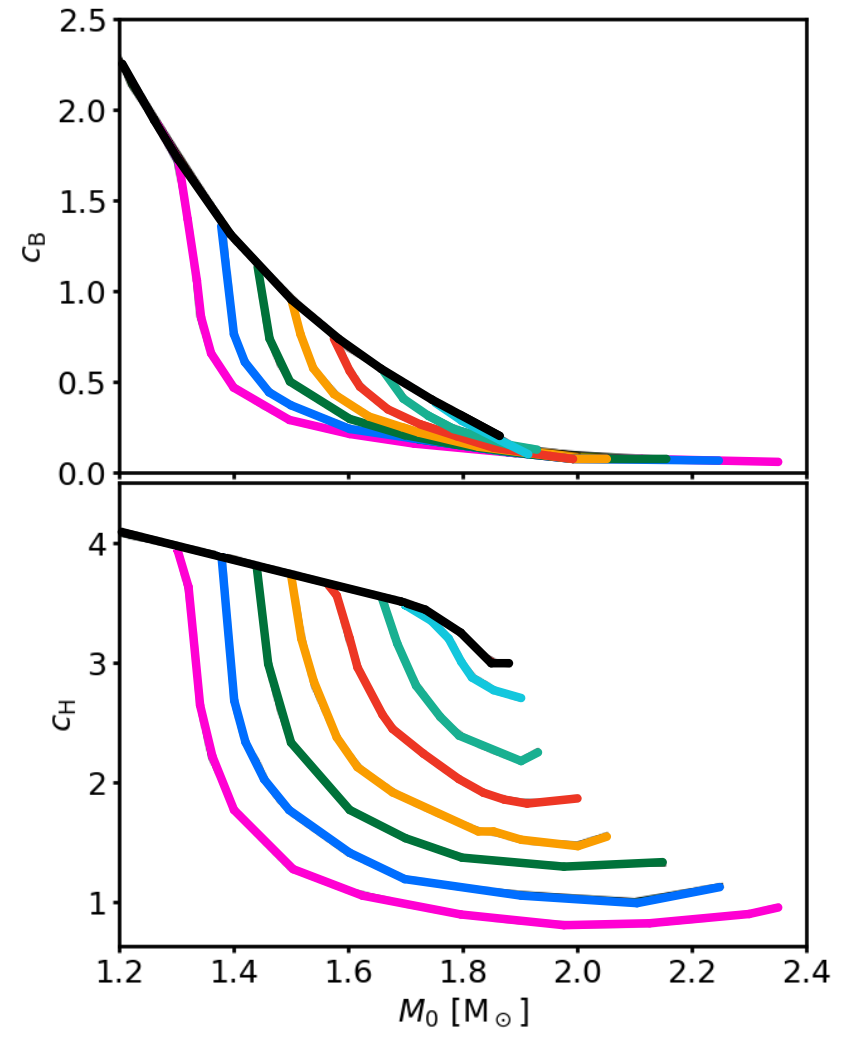}
    \caption{Distortion coefficients $c_\mathrm{B}$ (top panels) and $c_\mathrm{H}$ (bottom panels) as functions of the baryonic mass $M_0$ of models with a purely toroidal magnetic field (left panels) and with a purely poloidal magnetic field (right panels), for various value of $\beta _0$: from $\beta _0=-6$ (magenta curve)  to $\beta _0=-4.5$ (light blue curve) increasing by 0.25 with every line. The black curve corresponds to GR. Adapted from Soldateschi et al., A\&A, 645, A39 (2021).}
    \label{fig:cbch}
\end{figure*}

The coefficients $c_\mathrm{B}$ and $c_\mathrm{H}$ as functions of $M_0$, for various $\beta _0$, are shown in Fig.~\ref{fig:cbch}, in the case of NSs endowed with a purely toroidal or a purely poloidal magnetic field and described by the POL2 EoS $\tilde{p}=K_{\mathrm a} \tilde{\rho}^{\gamma_{\mathrm a}}$, with $\gamma_{\mathrm a}=2$ and $K_{\mathrm a}=110$ in dimensionless units. The black line represents GR, which corresponds to $\alpha_0 = \beta_0 = 0$. The other lines stand for $\beta _0 = \{ -4.5,-4.75,-5.0,-5.25,-5.5,-5.75,-6.0 \}$ and $\alpha _0 = -2 \times 10^{-4}$. The effect of scalarisation is visible in the distinctive change in the slope as the scalarised sequences depart from the GR one.  The more negative $\beta _0$ is, the more enhanced are the modifications with respect to GR. At a fixed $M_0$, scalarised NSs have lower distortion coefficient and quadrupole deformation than the GR models with the same mass for most of the scalarisation range. We found that, in the the ranges $1.2 \leq M_0/\mathrm{M}_\odot \lesssim 2.4 $ and $-6 \leq \beta _0 \leq -4.5 $, the distortion coefficients are well approximated by a combination of power laws of the baryonic mass $M_0$, the J-frame circumferential radius $\tilde{R}_\mathrm{c}$ and the E-frame scalar charge $\bar{Q}_\mathrm{s}$\citep{soldateschi_2020}. In particular
\begin{equation}\label{eq:distcoeff}
	c_\mathrm{B}\approx c_1 M_{0;1.6}^\alpha R_{10}^\beta \left[ 1-c_2 Q_{1}^\gamma M_{1.6}^\delta R_{10}^\rho  \right] \; ,
\end{equation}
where $M_{0;1.6}$ is $M_0$ in units of $1.6 \mathrm{M}_\odot$, $R_{10}$ is $\tilde R _\mathrm{c}$ in units of $10\mathrm{km}$, $Q_{1}$ is $\bar Q _\mathrm{s}$ in units of $1 \mathrm{M}_\odot$, $[c_1,\alpha,\beta,c_2,\gamma,\delta,\rho]=[0.16,-2.22,4.86,0.87,1.32,-1.27,-2.21]$ in the toroidal case and $[c_1,\alpha,\beta,c_2,\gamma,\delta,\rho]=[0.077,-1.99,5.80,1.38,1.22,-0.86,-3.49]$ in the poloidal case.
The coefficient $c_\mathrm{H}$ in the toroidal cases is similar to that in the poloidal case; in fact, it is possible to find a functional form over the entire mass range that holds for both magnetic configuartions with an accuracy of few percents:
\begin{equation}\label{eq:chapprox}
	c_\mathrm{H}\approx 0.5+ \mathcal{F}(M_0)\mathcal{T}(M_0,\bar Q _\mathrm{s},\tilde R _\mathrm{c})\times\begin{cases} 0.65\;{\rm for\; toroidal \;,} \\
	1.02\; {\rm for\; poloidal \;,}\end{cases} \\
	%(4.95-1.82 M_{1.6})\left\{1-\frac{1.8}{R_{10}^{2.34}}\left(\frac{Q_1}{M_{1.6}}\right)^{1.2} \right\},
\end{equation}
where $\mathcal{F}(M_0)$ does not depend explicitly on the scalar field and described the role of the equation of state, $\mathcal{T}(M_0,\bar Q _\mathrm{s},\tilde R _\mathrm{c})$ represents the correction due to the presence of scalarisation, and the last numerical factor differentiates the oblate from the prolate geometry, which depends on the magnetic field configuration. We found that:
\begin{eqnarray}\label{eq:chterms}
&\mathcal{F}(M_0) &= 4.98-1.95 M_{0;1.6} \;, \\
&\mathcal{T}(M_0,\bar{Q}_\mathrm{s},\tilde{R}_\mathrm{c})&=1-\frac{1.90}{R_{10}^{2.45}}\left(\frac{Q_1}{M_{0;1.6}}\right)^{1.3} \;.
\end{eqnarray}
%%%%%%%%%%%%%%%%%%% FIG 1 %%%%%%%%%%%%%%%%%%%%%%%

%%%%%%%%%%%%%%%%%%%%%%%%%%%%%%%%%%%%%%%%%%%%%%%%%%%%

\section{Selection of equations of state}\label{sec:6}
The results shown in Sect.~\ref{sec:4} are found using a simple polytropic law which, while allowing us to compare our results to previous studies \citep{frieben_equilibrium_2012,pili_axisymmetric_2014}, is not allowed by observations. For this reason, we further computed the distortion coefficients for a selection of EoS allowed by observational and nuclear physics constraints \citep{soldateschi_2021_1}. In particular, we selected 13 different EoS that span a diverse range of calculation methods and particle contents. All EoS we used, except the polytropic one, are chosen to satisfy the lates constraints: they reach a maximum mass of at least $\sim$2.05M$_\odot$\citep{Fonseca_2021}, satisfy various nuclear physics constraints \citep{Fortin_Providencia_Raduta_Gulminelli_Zdunik_Haensel_Bejger_2016}, are not too stiff \citep{chaves_2019}, and give the NS a radius between $\sim$10km and $\sim$14km for 1.4M$_\odot$ mass models {\citep{bauswein_2017,riley_2021}}. We chose 6 EoS which contain only $n p e \mu$ particles (APR,SLY9,BL2,DDME2,NL3$\omega \rho$,SFH), 2 EoS containing also hyperons (DDME2-Y,NL3$\omega \rho$-Y), 2 EoS containing an $u d s$ quark matter domain treated with the Nambu-Jona-Lasinio model (BH8, BF9), 2 EoS containing an $u d s$ quark matter domain treated with the MIT bag model or perturbative QCD (SQM1, SQM2), as well as the POL2 EoS.

\begin{figure}
   	\centering
         \includegraphics[width=0.503\textwidth]{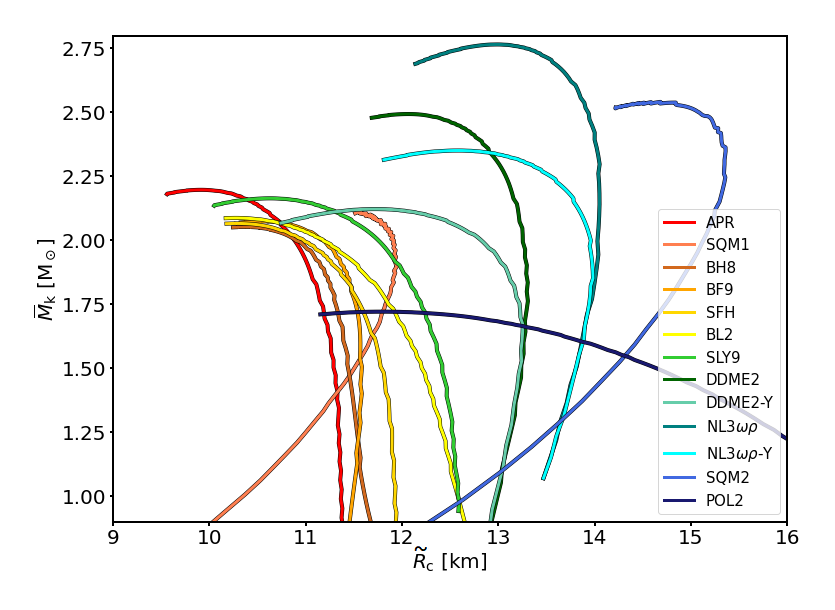}
         \includegraphics[width=0.487\textwidth]{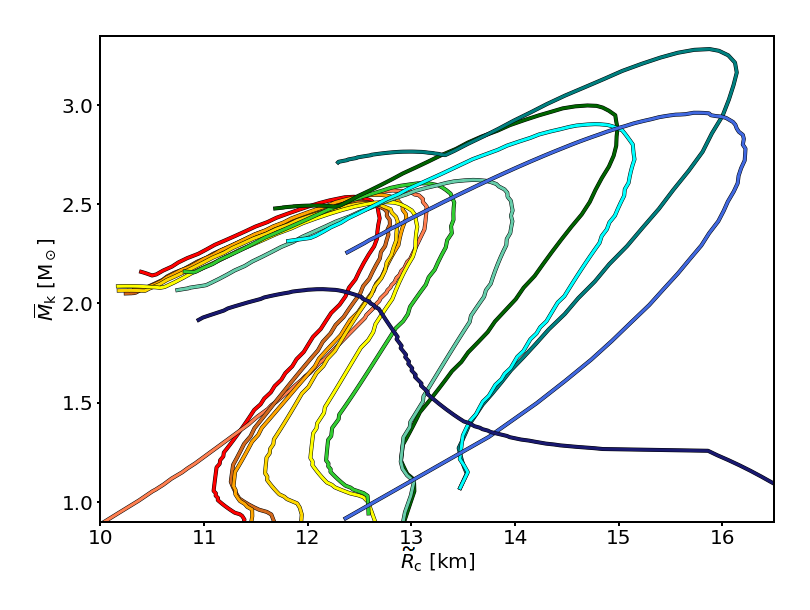}
         \caption{Komar mass $\bar M_\mathrm{k}$ against circumferential radius $\tilde R_\mathrm{c}$ for un-magnetised, static models of NSs computed with the EoS described in Sect.~\ref{sec:6} in GR (left plot) and in STT with $\beta_0=-6$ (right plot). The EoS are color-coded, and ordered in the legend, according to the compactness $C=M_\mathrm{k}/R_\mathrm{c}$ calculated at $M_\mathrm{k}=1.4$M$_\odot$ in GR: red for the highest compactness and blue for the lowest compactness. Adapted from Soldateschi et al., ArXiv e-prints (2021), arXiv:2106.00603v2 [astro-ph.HE].}
         \label{fig:mrrel}
 \end{figure}

In Fig.~\ref{fig:mrrel} we plot the Komar mass $\bar M_\mathrm{k}$ against the circumferential radius $\tilde R_\mathrm{c}$ for models of un-magnetised, static NSs computed with the described EoS. The left panel refers to GR, while the right panel to STT with $\beta_0=-6$. From the left panel of Fig.~\ref{fig:mrrel} we see that the NS radii have values ranging from $\sim$10km to $\sim$14km for most EoS, excluding the less compact SQM2 and especially the POL2 EoS. The maximum masses are around $\sim$2-2.2M$_\odot$ for most EoS, except for the NL3$\omega \rho$ and the POL2 EoS.
%%%%%%%%%%%%%%%%%%%%%%%%% FIG 2 %%%%%%%%%%%%%%%%%%%%%%%%%
%%%%%%%%%%%%%%%%%%%%%%%%%%%%%%%%%%%%%%%%%%%%%%%%%%%%%%%%%

\section{Quasi-universal relations for the magnetic deformation of neutron stars}\label{sec:7}
We computed the distortion coefficients in Eq.~\ref{eq:coeffs} for the EoS described in Sect.~\ref{sec:6}, as well as another distortion coefficient:
\begin{equation}\label{eq:distcoeff_cs}
    |\bar e| = c_\mathrm{s} \tilde B^2_\mathrm{s} + \mathcal{O}\left( \tilde B^4_\mathrm{s} \right) \; ,
\end{equation}
where $\tilde B_\mathrm{s}$ is the magnetic field calculated at the pole of the NS, at the surface, normalised to $10^{18}$G. The coefficients $c_\mathrm{B}$ and $c_\mathrm{H}$ contain quantities that are not directly accessible by observations, because they require to know the details of the internal structure and magnetic field geometry of NSs. On the other hand, $c_\mathrm{s}$ may prove to be more useful to compare our results to observations. This last coefficient is defined only for configurations endowed with a poloidal magnetic field (the toroidal one being hidden under the surface).

\begin{figure}[!htp]
  \centering
  \includegraphics[width=0.49\textwidth]{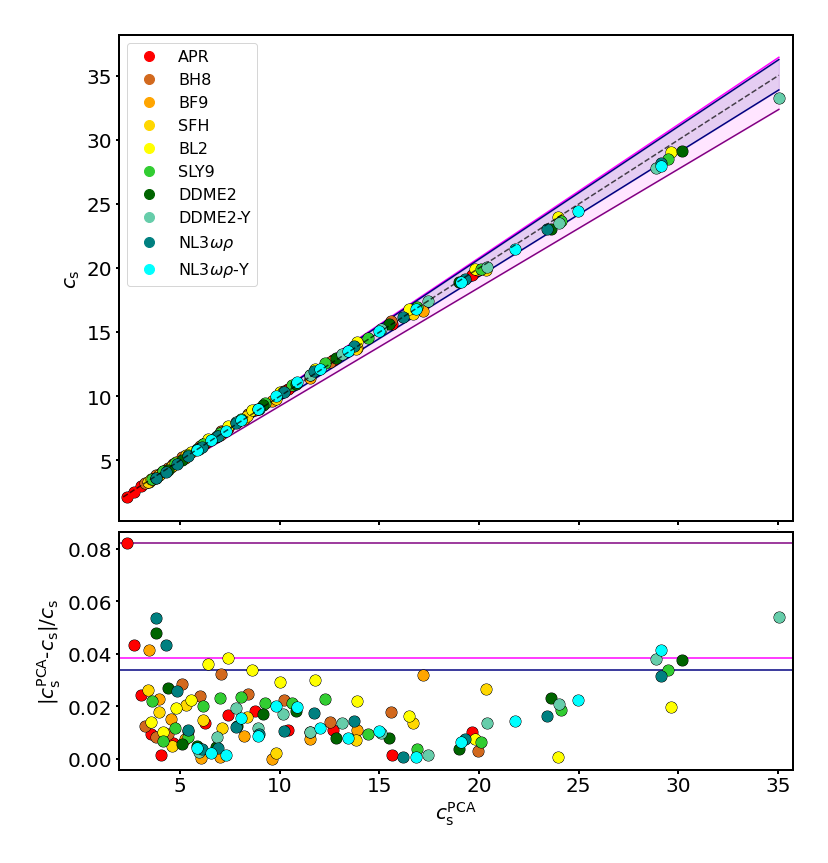}
  \includegraphics[width=0.49\textwidth]{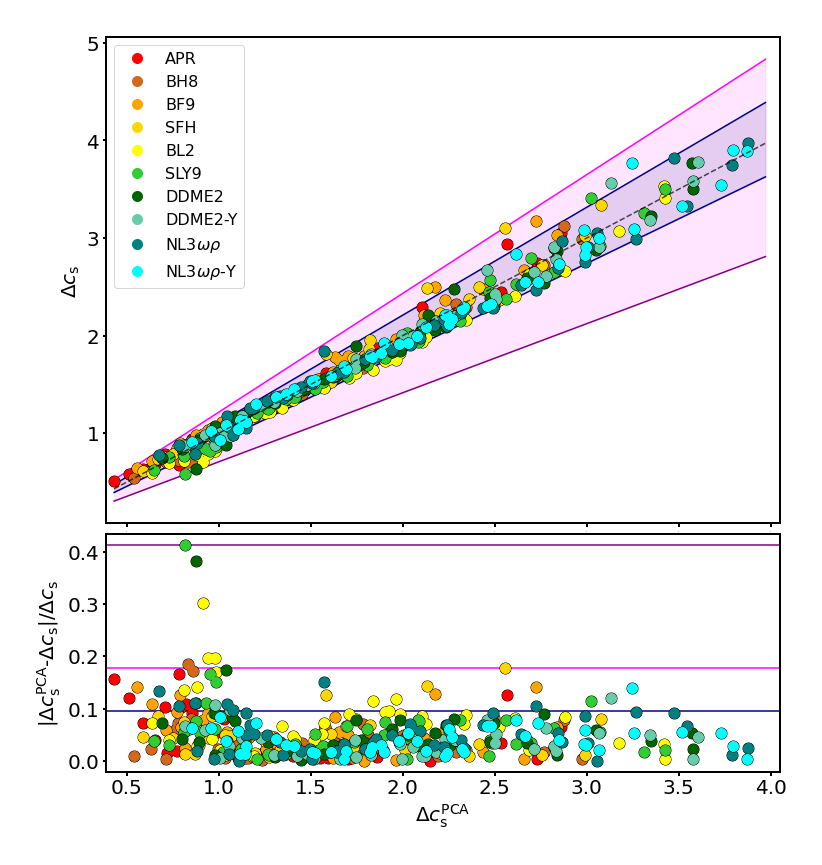}
  \caption{
Top left plot: distortion coefficient $c_\mathrm{s}$, calculated according to Eq.~\ref{eq:distcoeff_cs} in GR, versus its approximation $c^\mathrm{PCA}_\mathrm{s}$ calculated with the quasi-universal relation in Eq.~\ref{eq:cs_pca_gr}.
Top right plot: Difference $\Delta c_\mathrm{s}$ between the distortion coefficient $c_\mathrm{s}$, calculated according to Eq.~\ref{eq:distcoeff_cs} in STT with $\beta _0~\in~\{ -6,-5.75,-5.5,-5 \}$, and the GR quasi-universal relation in Eq.~\ref{eq:cs_pca_gr}. This is plotted versus its approximation $\Delta c^\mathrm{PCA}_\mathrm{s}$, calculated with the quasi-universal relation in Eq.~\ref{eq:cs_pca_stt}.
The corresponding relative deviations from the PCA are given in the bottom plots.
The dashed lines are $c_\mathrm{B,H} = c^\mathrm{PCA}_\mathrm{B,H}$ and $\Delta c_\mathrm{s} = \Delta c^\mathrm{PCA}_\mathrm{s}$, respectively.
The magenta shaded areas comprise all data points and the purple and magenta lines represent the upper and lower bounds of Eq.~\ref{eq:cs_pca_gr} and Eq.~\ref{eq:cs_pca_stt}, respectively; the dark blue lines bounding the shaded blue area mark the 90th percentile error region.
The EoS are color-coded, and ordered in the legend, according to the compactness $C=M_\mathrm{k}/R_\mathrm{c}$ calculated at $M_\mathrm{k}=1.4$M$_\odot$ in GR: red for the highest compactness and blue for the lowest compactness.
Adapted from Soldateschi et al., ArXiv e-prints (2021), arXiv:2106.00603v2 [astro-ph.HE].}
  \label{fig:pca_cs}
\end{figure}

We found that $c_\mathrm{B},c_\mathrm{H}$ and $c_\mathrm{s}$ are similar for all standard EoS (i.e. all EoS except SQM1, SQM2 and POL2) as functions of the Komar mass \citep{soldateschi_2021_1}. For this reason, we chose to consider the dependence also on another potentially observable quantity, namely the circumferential radius $\tilde R_\mathrm{c}$, to understand whether the spread among the various EoS could be further reduced. We used the `principal component analysis' (PCA) technique to find the best-fit relation between $c_\mathrm{B,H,s}$, $\bar M_\mathrm{k}$ and $\tilde R_\mathrm{c}$. We found the following formulas approximating $c_\mathrm{B,H}$ in GR to a satisfying level of accuracy for all standard EoS (thus the name `quasi-universal relations'):
\begin{equation}\label{eq:cb_pca_gr}
        c^\mathrm{PCA}_\mathrm{B} =
            \begin{cases}
                0.13^{+0.03}_{-0.02} R_{10}^{5.45} M_{1.6}^{-2.41} \;{\rm for\; poloidal,}\\\\
                0.25^{+0.03}_{-0.03} R_{10}^{5.03} M_{1.6}^{-2.07} \;{\rm for\; toroidal,}
            \end{cases}
\end{equation}
\begin{align}\label{eq:ch_pca_gr}
        c^\mathrm{PCA}_\mathrm{H} =
            \begin{cases}
                5.77^{+0.04}_{-0.06} - 0.77 R_{10} - 4.14 M_{1.6} - 0.27 M_{1.6}^2 +\\ \;\;+ 0.07 R_{10}^2 + 2.28 M_{1.6} R_{10} \;{\rm for\; poloidal,}\\\\
                7.02^{+0.05}_{-0.07} - 5.22 R_{10} - 2.76 M_{1.6} - 0.12 M_{1.6}^2 +\\ \;\;+ 1.92 R_{10}^2 + 1.51 M_{1.6} R_{10} \;{\rm for\; toroidal,}
            \end{cases}
\end{align}
where $M_{1.6} = \bar M_\mathrm{k}/1.6$M$_\odot$. Moreover, we found that, in GR, $c_\mathrm{s}$ is well-approximated by the following relation:
\begin{equation}\label{eq:cs_pca_gr}
    \begin{split}
        c^\mathrm{PCA}_\mathrm{s} &= 2.97^{+0.12}_{-0.23} R_{10}^{4.61} M_{1.6}^{-2.80} \; .
    \end{split}
\end{equation}
The coefficient $c_\mathrm{s}$ computed using Eq.~\ref{eq:cs_pca_gr} is plotted against the value computed with formula Eq.~\ref{eq:distcoeff_cs} in Fig.~\ref{fig:pca_cs} (top left plot), together with the error of the approximation (bottom left plot). The dashed black line is a reference $c_\mathrm{s} = c^\mathrm{PCA}_\mathrm{s}$ bisecting line, which represents a perfect approximation. The superscript and subscript in the first coefficient of Eq.~\ref{eq:cs_pca_gr} are the values defining the purple and magenta lines bounding the magenta shaded area in Fig.~\ref{fig:pca_cs}, top left plot. The dark blue line, bounding the shaded blue area, marks the 90th percentile of the errors (the bounds containing 90\% of the results). The corresponding values of these errors are showed with lines of the same colour in the bottom left plot. The approximation $c^\mathrm{PCA}_\mathrm{s}$ holds with a very small error, under a few percents. Moreover, we found that the quasi-universal relations for $c_\mathrm{B}$ hold with a slightly larger error, around $\sim 10\%$. The approximation for $c_\mathrm{H}$ is even more accurate, with an error that is mostly under $\sim 1\%$. Moreover, we found that performing $c^\mathrm{PCA}_\mathrm{H} \rightarrow 5/3c^\mathrm{PCA}_\mathrm{H}-0.9$ allows one to use the coefficients found in the toroidal case also in the poloidal configurations with a $\sim 2\%$ error.

Applying the quasi-universal relations derived in the case of standard EoS, Eq.s~\ref{eq:cb_pca_gr}-\ref{eq:ch_pca_gr}-\ref{eq:cs_pca_gr}, to the polytropic models computed with the POL2 EoS, leads to larger errors, especially in the case of $c_\mathrm{B}$. We refer the reader to Ref.~\citenum{soldateschi_2021_1} for more details.
\\\\
In the case of STTs, we find quasi-universal relations for $\Delta c_\mathrm{B} = |c_\mathrm{B} - c^\mathrm{GR}_\mathrm{B}|$, $\Delta c_\mathrm{H} = |c_\mathrm{H} - c^\mathrm{GR}_\mathrm{H}|$ and $\Delta c_\mathrm{s} = |c_\mathrm{s} - c^\mathrm{GR}_\mathrm{s}|$, where $c^\mathrm{GR}_\mathrm{B}, c^\mathrm{GR}_\mathrm{H}$ and $c^\mathrm{GR}_\mathrm{s}$ are the relations found in the GR case: Eq.s~\ref{eq:cb_pca_gr}-\ref{eq:ch_pca_gr}-\ref{eq:cs_pca_gr} respectively. In this case, we also allowed the dependence of the distortion coefficients on the scalar charge $\bar Q_\mathrm{s}$. We found the following quasi-universal relations:
\begin{equation}\label{eq:cb_pca_stt}
        \Delta c^\mathrm{PCA}_\mathrm{B} =
            \begin{cases}
                0.03^{+0.05}_{-0.03} R_{10}^{8.23} M_{1.6}^{-5.08} Q_{1}^{2.60} \;{\rm for\; poloidal,}\\\\
                0.06^{+0.09}_{-0.05} R_{10}^{5.96} M_{1.6}^{-3.52} Q_{1}^{1.95} \;{\rm for\; toroidal,}
            \end{cases}
\end{equation}
\begin{equation}\label{eq:ch_pca_stt}
        \Delta c^\mathrm{PCA}_\mathrm{H} =
            \begin{cases}
                1.96^{+0.17}_{-0.18} R_{10}^{0.72} M_{1.6}^{-1.96} Q_{1}^{1.54} \;{\rm for\; poloidal,}\\\\
                1.49^{+0.26}_{-0.17} R_{10}^{0.75} M_{1.6}^{-1.81} Q_{1}^{1.55} \;{\rm for\; toroidal,}
            \end{cases}
\end{equation}
\begin{equation}\label{eq:cs_pca_stt}
        \Delta c^\mathrm{PCA}_\mathrm{s} =
                0.92^{+0.20}_{-0.27} R_{10}^{4.77} M_{1.6}^{-4.50} Q_{1}^{1.71} \;.
\end{equation}

The quasi-universal relation in Eq.~\ref{eq:cs_pca_stt} is plotted against the corresponding value $\Delta c_\mathrm{s}$, computed through formula Eq.~\ref{eq:distcoeff_cs}, in Fig.~\ref{fig:pca_cs} (top right plot). The bottom right plot displays the relative error of the quasi-universal relation. The dashed line is a reference $\Delta c_\mathrm{s} = \Delta c^\mathrm{PCA}_\mathrm{s}$ bisecting line, which would be a perfect approximation. We can see that the approximation for $c_\mathrm{s}$ holds with an error that is around $10\%$. Moreover, we found that the quasi-universal relations for $\Delta c_\mathrm{B}$ hold with an error of $\sim 50\%$. The approximation for $\Delta c_\mathrm{H}$ is more accurate, with a relative error of just a few percents. Similarly to what we found in the GR case, performing $c^\mathrm{PCA}_\mathrm{H} \rightarrow 3/2c^\mathrm{PCA}_\mathrm{H}-0.2$ allows us to use the coefficients found in the toroidal case in the poloidal geometry with a $\sim 10\%$ error.

\begin{figure*}[!ht]
  \centering
  \includegraphics[width=.49\textwidth]{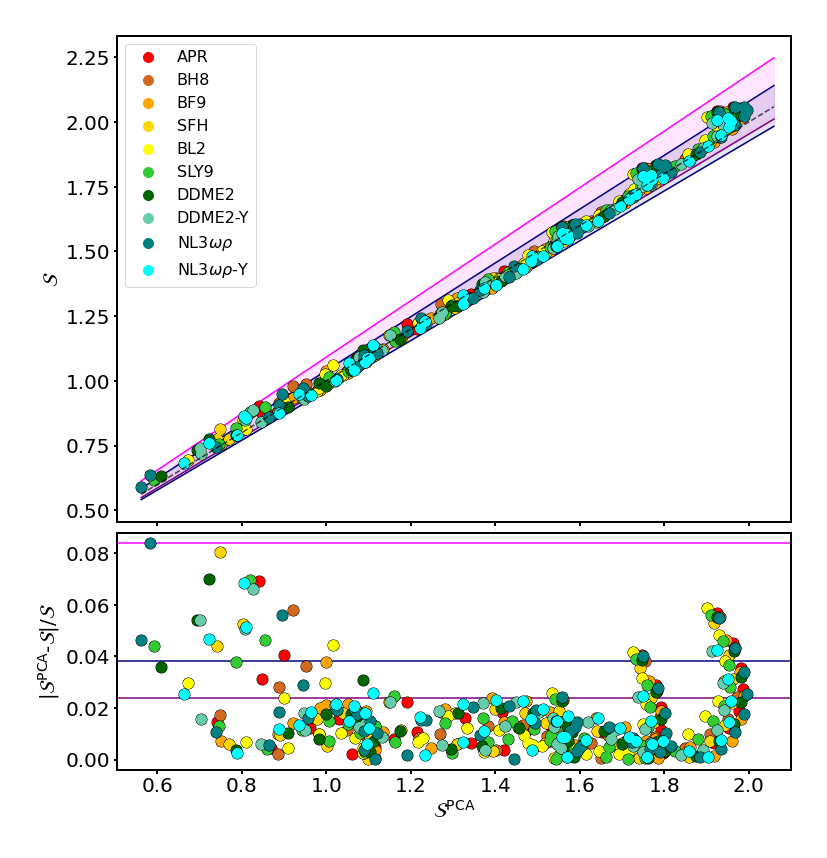}
  \includegraphics[width=.49\textwidth]{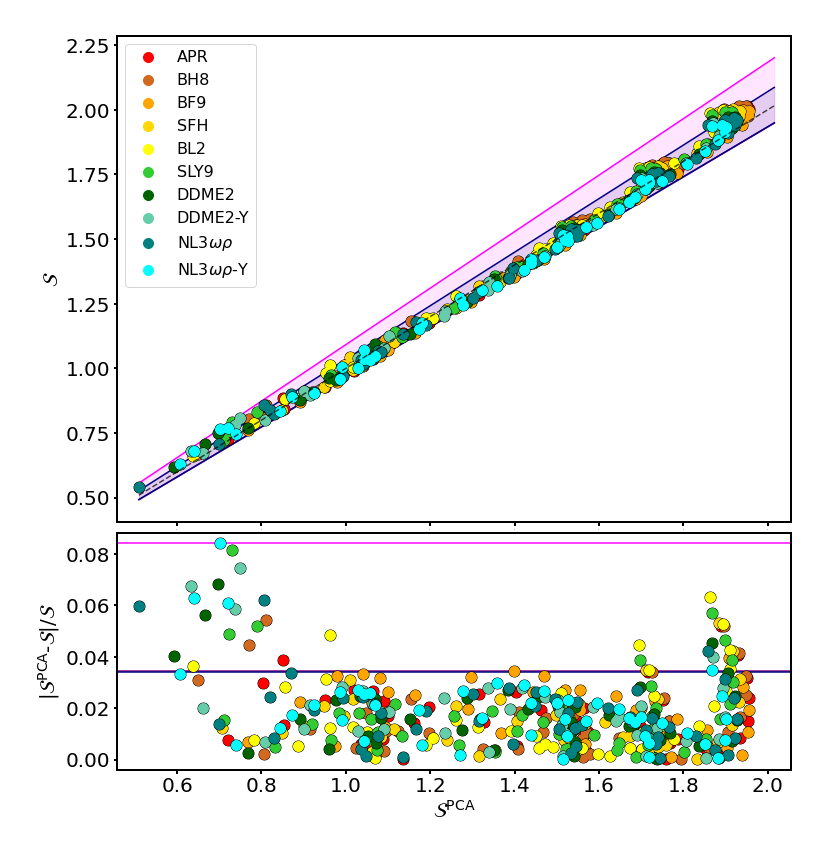}
  \caption{Ratio $\mathcal{S}$ between scalar and tensor quadrupolar GW losses, calculated according to Eq.~\ref{eq:sratio} in STT with $\beta _0~\in~\{ -6,-5.75,-5.5,-5 \}$. This is plotted versus its approximation $\mathcal{S}^\mathrm{PCA}$, calculated with the quasi-universal relations in Eq.~\ref{eq:sratio_pca_stt} (top plot in each panel). The corresponding relative devitations from the PCA are given in the bottom plot in each panel. The left panel refers to a purely poloidal magnetic field; the right panel refers to a purely toroidal magnetic field. The dashed line is $\mathcal{S} = \mathcal{S}^\mathrm{PCA}$. The magenta shaded area comprises all data points and the purple and magenta lines represent the upper and lower bounds of Eq.~\ref{eq:sratio_pca_stt}; the dark blue lines bounding the shaded blue area mark the 90th percentile error region. The EoS are color-coded, and ordered in the legend, according to the compactness $C=M_\mathrm{k}/R_\mathrm{c}$ calculated at $M_\mathrm{k}=1.4$M$_\odot$ in GR: red for the highest compactness and blue for the lowest compactness. Adapted from Soldateschi et al., ArXiv e-prints (2021), arXiv:2106.00603v2 [astro-ph.HE].}
  \label{fig:pca_sratio_stt}
\end{figure*}

Using Eq.~\ref{eq:cb_pca_stt} for models computed using the POL2 EoS, errors in approximating $\Delta c_\mathrm{B}$ remain of the same magnitude in the poloidal case, while they increase by $\sim 20\%$ in the toroidal case. Instead, Eq.~\ref{eq:ch_pca_stt} holds also for the POL2 EoS, at the expense of an error reaching a few tens of percents for the approximation of $\Delta c_\mathrm{H}$, as in the case of approximating $\Delta c_\mathrm{s}$ using Eq.~\ref{eq:cs_pca_stt} for polytropic models computed using the POL2 EoS.
%%%%%%%%%%%%%%%%%%%%%%%%% FIG 3 %%%%%%%%%%%%%%%%%%%%%%%%%
%%%%%%%%%%%%%%%%%%%%%%%%%%%%%%%%%%%%%%%%%%%%%%%%%%%%%%%%%
\\\\
As in the case of the POL2 EoS, we incur in larger errors if we apply the quasi-universal relations we found to the case of the SQM1 and SQM2 EoS. In GR, $c^\mathrm{PCA}_\mathrm{B}$ is a factor $\sim 0.4-0.8$ lower than $c_\mathrm{B}$. The maximum error for approximating $c_\mathrm{H}$ with Eq.~\ref{eq:ch_pca_gr} increases to $\sim 8-12\%$ in the case of SQM1, and $\sim 5\% (\sim 40\%)$ for purely poloidal (toroidal) magnetic fields in the case of SQM2. Moreover, $c^\mathrm{PCA}_\mathrm{s}$ is at most a factor $\sim 1.4$ higher than $c_\mathrm{s}$. In STT, $\Delta c^\mathrm{PCA}_\mathrm{B}$ in the poloidal case is around a factor $\sim 2$ lower (higher) than $\Delta c_\mathrm{B}$ for the SQM1 (SQM2) EoS; in the toroidal case, it is a factor $\sim 2$ lower for both SQM1 and SQM2. The maximum error for approximating $\Delta c_\mathrm{H}$ using Eq.~\ref{eq:ch_pca_stt} increases to $\sim 30\% (\sim 50\%)$ for purely poloidal (toroidal) magnetic fields. Finally, $\Delta c^\mathrm{PCA}_\mathrm{s}$ is around a factor $\sim 1.7$ lower than $\Delta c_\mathrm{s}$.
\\\\
A time-varying quadrupolar deformation leads to the emission of GWs. While in GR these are only of tensor nature, in the case of STTs a scalar channel is also present, which can contain any multipolar component. We only focus here on quadrupolar modes of GWs, both of tensor and scalar nature. As we have shown, NSs in GR and in STTs posses quite different quadrupolar deformations; for this reason, we compare the amount of GWs emitted by NSs in these two modes. To this end, we introduce the following quantity:
\begin{equation}\label{eq:sratio}
        \mathcal{S}= \bigg | \frac{q_\mathrm{s}}{q_\mathrm{g}} \bigg |,
\end{equation}
where
\begin{align}
        &q_\mathrm{s} = 2\pi \int \alpha _\mathrm{s} \mathcal{A}^4 \tilde T \left( 3\sin ^2 \theta -2 \right) r^4 \sin \theta \mathrm{d}r \mathrm{d}\theta , \\
        &q_\mathrm{g} = \int \left[ \pi \mathcal{A}^4 (\tilde \varepsilon+\tilde \rho)-\frac{1}{8} \partial \chi \partial \chi \right] r^4 \sin \theta \left(3\sin^2\theta -2 \right)  \mathrm{d}r \mathrm{d}\theta .
\end{align}
The quantity $q_\mathrm{s}$ is related to the source of scalar waves. The mass quadrupole $q_\mathrm{g}$ is $\bar I_\mathrm{zz}- \bar I_\mathrm{xx}=\bar e \bar I_\mathrm{zz}$ and is the source of tensor waves. Thus, the ratio $\mathcal{S}$ computed for a given NS model measures the relative amount of quadrupolar GWs emitted in the scalar and tensor channels. If $\mathcal{S}<1$, the majority of GWs will be emitted in the tensor channel; if $\mathcal{S}>1$, in the scalar channel. We found that the following quasi-universal relations hold for $\mathcal{S}$:
\begin{equation}\label{eq:sratio_pca_stt}
        \mathcal{S}^\mathrm{PCA}=
            \begin{cases}
                1.98^{+0.18}_{-0.05} R_{10}^{-0.71} M_{1.6}^{-0.54} Q_{1}^{1.22} \;{\rm for\; poloidal,}\\\\
                1.99^{+0.18}_{-0.07} R_{10}^{-0.74} M_{1.6}^{-0.60} Q_{1}^{1.23} \;{\rm for\; toroidal.}
            \end{cases}
\end{equation}
In Fig.~\ref{fig:pca_sratio_stt} we plot the values of $\mathcal{S}$, computed through Eq.~\ref{eq:sratio} for $\beta _0~\in~\{-6,-5.75,-5.5,-5\}$, against their PCA approximation, computed through Eq.~\ref{eq:sratio_pca_stt}. The approximation is very accurate in both magnetic geometries, with an error that is mostly concentrated under $\sim 4\%$. Moreover, we see that the coefficients in Eq.~\ref{eq:sratio_pca_stt} are almost identical in the poloidal and the toroidal cases, thus the two approximations are practically equivalent. It is possible that this similarity means that there exist a relation between the sources of scalar and tensor waves, $q_\mathrm{s}$ and $q_\mathrm{g}$, that does not depend on either the magnetic field geometry or the EoS.
%%%%%%%%%%%%%%%%%%%%%%%%% FIG 4 %%%%%%%%%%%%%%%%%%%%%%%%%
%%%%%%%%%%%%%%%%%%%%%%%%%%%%%%%%%%%%%%%%%%%%%%%%%%%%%%%%%
\\\\
The quasi-universal relations we found are independent on the EoS of standard NSs. For this reason they may be useful in disentangling the effects of the EoS and the magnetic field structure of NSs, leaving their internal magnetic structure as the only major unknown in GR. In this sense, the most promising relation is that of $c_\mathrm{s}$. On the one hand, $c_\mathrm{s}$ can be computed from its definition Eq.~\ref{eq:distcoeff_cs}. In this case, one needs to measure both the magnetic field strength at the surface of the NS, $B_\mathrm{s}$, and its quadrupolar deformation $e$, which, in turn, can be estimated from the strain of CGWs emitted by the NS: $h_0\propto eI$, where $I$ is the moment of inertia of the NS along its rotation axis, which can be computed through an EoS-independent relation by knowing the NS mass and its radius \citep{breu_2016}. On the other hand, our quasi-universal relation Eq.~\ref{eq:cs_pca_gr} allows one to estimate $c^\mathrm{PCA}_\mathrm{s}$ by knowing just the NS mass and radius. Comparing these values of $c_\mathrm{s}$ and $c^\mathrm{PCA}_\mathrm{s}$ can give an insight into the internal magnetic geometry of the emitting NS: the quasi-universal relation Eq.~\ref{eq:cs_pca_gr} was found in the case of a purely poloidal field, that is an extreme configuration which exherts the maximum possible magnetic deformation on the NS. For example, a lower value of $c_\mathrm{s}$ inferred by observations may imply the existence of a toroidal component of the NS internal magnetic field, which counteracts the effect of the poloidal component and results in a lower magnetic deformation. In the opposite case, another source of deformation may be present other than the magnetic field.

The quasi-universal relations Eq.s~\ref{eq:cb_pca_gr}-\ref{eq:ch_pca_gr} may be more useful to constrain $B_\mathrm{max}$ or $\mathcal{H}/W$ themselves, being these quantities generally not observable. In particular, $B_\mathrm{max} \approx (e/c_\mathrm{B})^{1/2}$ and $\mathcal{H}/W \approx e/c_\mathrm{H}$. As in the case of $c_\mathrm{s}$, we expect $c_\mathrm{B,H}<c^\mathrm{PCA}_\mathrm{B,H}$. If the NS mass and radius are known, one can esitimate a lower bound for both $B_\mathrm{max}$ and $\mathcal{H}/W$ by using the quasi-universal relations found in the purely poloidal and purely toroidal case. Moreover, the quasi-universal relations for $c_\mathrm{B}$, $c_\mathrm{H}$ and $c_\mathrm{s}$ can be useful to easily compute the distortion coefficients from the mass and radius of a model, without having to fully simulate the NS model.

In the case of STTs, some of the effects of the EoS are degenerate with the presence of a scalar charge inside the NS. For this reason, quasi-universal relations Eq.s~\ref{eq:cb_pca_stt}-\ref{eq:ch_pca_stt}-\ref{eq:cs_pca_stt} may help to understand whether a distortion coefficient inferred through relations Eq.s~\ref{eq:distcoeff}-\ref{eq:distcoeff_cs}, is compatible with the observed NS possessing a scalar charge, independently from its EoS. Relation Eq.~\ref{eq:sratio_pca_stt} does not involve knowing the strength of the magnetic field, thus it is probably more promising. In particular, following our previous argument, it may be that $\mathcal{S} \approx \mathcal{S}^\mathrm{PCA}$ independently of the magnetic configuration. Then, the observation of CGWs coming from a given NS of known mass, radius, distance $d$ and spin period $P$ translates into a lower bound for a function of the scalar charge. On the other hand, the non-observation of CGWs from a given NS can be translated into an upper bound.

\begin{figure*}[!ht]
  \centering
  \includegraphics[width=\textwidth]{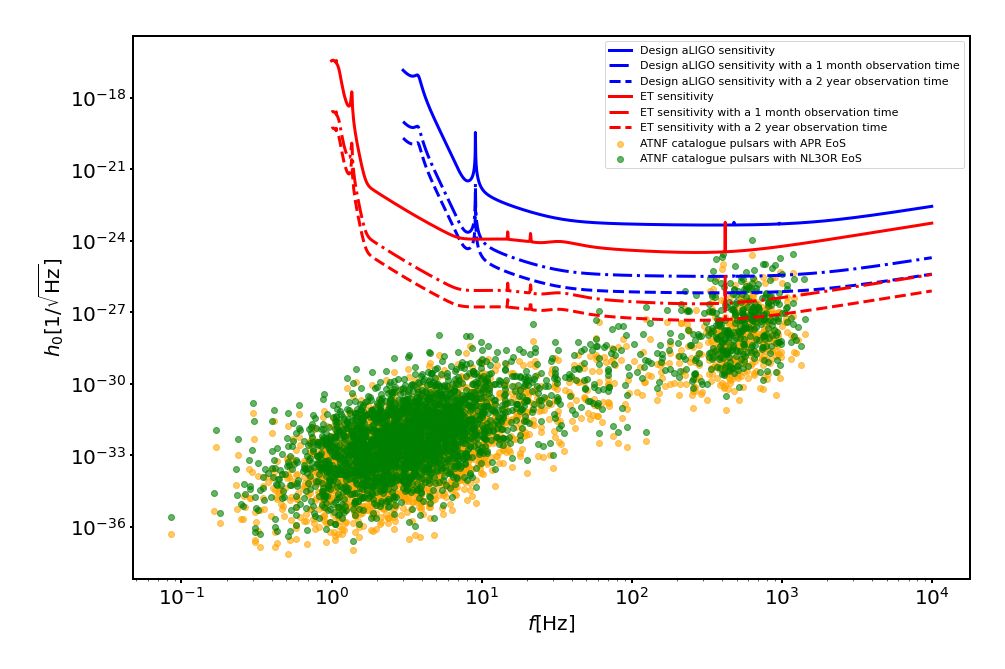}
  \caption{CGWs strain $h_0$ computed for pulsars contained in the ATNF catalogue through the use of quasi-universal relation Eq.~\ref{eq:cs_pca_gr} and by assuming the APR EoS (orange points) or the NL3$\omega \rho$ EoS (green points). The x-axis represents the frequency $f$ of the emitted CGWs. The blue (red) solid line represents the sensitivity of the aLIGO (ET) detector, while the blue (red) dot-dashed and dashed lines show the minimum strain detectable by aLIGO (ET) assuming a 1 month or 2 years observing time, respectively.}
  \label{fig:strain}
\end{figure*}

Finally, we can use the quasi-universal relations we found to assess the detectability of known NSs, in GR. Since the strain of CGWs emitted by a deformed NS, rotating with frequency $f_\mathrm{rot}$ and at a distance $d$, is $h_0 \propto eIf^2_\mathrm{rot}/d$, by using the quasi-universal relation for $c_\mathrm{s}$ Eq.~\ref{eq:cs_pca_gr} we can compute $h_0$ for the pulsars of the ATNF \citep{atnf_2005} catalogue. In this case, $f_\mathrm{rot}$ and $d$ are taken from the data in ATNF catalogue, while $M_\mathrm{k},B_\mathrm{s}$ are generated from the expected distributions \citep{antoniadis_2016,faucher_2006}. The radius $R_\mathrm{c}$ is computed by assuming the two most diverse standard EoS among the ones we considered, APR and NL3$\omega \rho$, and using the corresponding mass-radius diagram. Our results are shown in Fig.~\ref{fig:strain}, where they are compared to the sensitivity of the advanced LIGO (aLIGO) detector in the design configuration\footnote{The aLIGO design densitivity curves can be found at \url{https://dcc.ligo.org/LIGO-T1800044/public}.} and of the Einstein Telescope (ET) detector in the D configuration\footnote{The ET sensitivity curves can be found at \url{http://www.et-gw.eu/index.php/etsensitivities}.}: the blue (red) solid line represents the sensitivity of aLIGO (ET), while the blue (red) dot-dashed and dashed lines show the minimum strain detectable by aLIGO (ET) assuming a 1 month or 2 years observing time, respectively. The x-axis represents the frequency $f$ of the emitted CGWs. In the general case, CGWs are emitted at two frequencies, $f=f_\mathrm{rot}$ and $f=2f_\mathrm{rot}$. In this case, we consider only the $f=2f_\mathrm{rot}$ wave. As we can see from Fig.~\ref{fig:strain}, many millisecond pulsars contained in the ATNF catalogue have a chance to emit CGWs which could be potentially observed with the future ET detector, especially considering a 1 month or 2 year observing campaign. As for aLIGO, in the case of 1 month or 2 years observing time, a few tens of the ATNF millisecond pulsars could potentially be detected through CGWs. We note that the difference in radii given by the two different EoS does not significantly alter the strain of the emitted CGWs. However, pulsars with smaller rotation period are much less likely to be observed, even with a 3rd generation detector like ET.
%%%%%%%%%%%%%%%%%%%%%%%%% FIG 5 %%%%%%%%%%%%%%%%%%%%%%%%%
%%%%%%%%%%%%%%%%%%%%%%%%%%%%%%%%%%%%%%%%%%%%%%%%%%%%%%%%%

\bibliographystyle{ws-procs961x669}
\bibliography{ws-pro-sample}

\begin{thebibliography}{10}

\bibitem{duncan_1992}
R.~C. {Duncan} and C.~{Thompson}, {Formation of Very Strongly Magnetized
  Neutron Stars: Implications for Gamma-Ray Bursts}, {\em \apjl} {\bf 392},
  p.~L9 (June 1992).

\bibitem{thompson_1993}
C.~{Thompson} and R.~C. {Duncan}, {Neutron Star Dynamos and the Origins of
  Pulsar Magnetism}, {\em \apj} {\bf 408}, p. 194 (May 1993).

\bibitem{thompson_1995}
C.~{Thompson} and R.~C. {Duncan}, {The soft gamma repeaters as very strongly
  magnetized neutron stars - I. Radiative mechanism for outbursts}, {\em
  \mnras} {\bf 275}, 255 (July 1995).

\bibitem{thompson_1996}
C.~{Thompson} and R.~C. {Duncan}, {The Soft Gamma Repeaters as Very Strongly
  Magnetized Neutron Stars. II. Quiescent Neutrino, X-Ray, and Alfven Wave
  Emission}, {\em \apj} {\bf 473}, p. 322 (December 1996).

\bibitem{asseo_2002}
E.~Asseo and D.~Khechinashvili, {The role of multipolar magnetic fields in
  pulsar magnetospheres}, {\em \mnras} {\bf 334}, 743 (08 2002).

\bibitem{spruit_source_2009}
H.~C. {Spruit}, {The source of magnetic fields in (neutron-) stars}, in {\em
  Cosmic Magnetic Fields: From Planets, to Stars and Galaxies\/},  eds. K.~G.
  {Strassmeier}, A.~G. {Kosovichev} and J.~E. {Beckman}, IAU Symposium,
  Vol.~259April 2009.

\bibitem{ferrario_magnetic_2015}
L.~Ferrario, A.~Melatos and J.~Zrake, Magnetic {Field} {Generation} in {Stars},
  {\em \ssr} {\bf 191}, p.~77 (October 2015).

\bibitem{olausen_2014}
S.~A. {Olausen} and V.~M. {Kaspi}, {The McGill Magnetar Catalog}, {\em \apjs}
  {\bf 212}, p.~6 (May 2014).

\bibitem{popov_origins_2016}
S.~B. {Popov}, {Origins of magnetars in binary systems}, {\em \aat} {\bf 29},
  183 (January 2016).

\bibitem{rea_2010}
N.~{Rea}, P.~{Esposito}, R.~{Turolla}, G.~L. {Israel}, S.~{Zane}, L.~{Stella},
  S.~{Mereghetti}, A.~{Tiengo}, D.~{G{\"o}tz}, E.~{G{\"o}{\u{g}}{\"u}{\c{s}}}
  and C.~{Kouveliotou}, {A Low-Magnetic-Field Soft Gamma Repeater}, {\em
  Science} {\bf 330}, p. 944 (November 2010).

\bibitem{staubert_2019}
R.~{Staubert}, J.~{Tr{\"u}mper}, E.~{Kendziorra}, D.~{Klochkov}, K.~{Postnov},
  P.~{Kretschmar}, K.~{Pottschmidt}, F.~{Haberl}, R.~E. {Rothschild},
  A.~{Santangelo}, J.~{Wilms}, I.~{Kreykenbohm} and F.~{F{\"u}rst}, {Cyclotron
  lines in highly magnetized neutron stars}, {\em \aap} {\bf 622}, p. A61
  (February 2019).

\bibitem{del_zanna_chiral_2018}
L.~Del~Zanna and N.~Bucciantini, {Covariant and 3 + 1 equations for
  dynamo-chiral general relativistic magnetohydrodynamics}, {\em \mnras} {\bf
  479}, 657 (06 2018).

\bibitem{ciolfi_2019}
R.~{Ciolfi}, W.~{Kastaun}, J.~V. {Kalinani} and B.~{Giacomazzo}, {First 100 ms
  of a long-lived magnetized neutron star formed in a binary neutron star
  merger}, {\em \prd} {\bf 100}, p. 023005 (July 2019).

\bibitem{franceschetti_2020}
K.~{Franceschetti} and L.~{Del Zanna}, {General Relativistic Mean-Field Dynamo
  Model for Proto-Neutron Stars}, {\em Universe} {\bf 6}, p.~83 (June 2020).

\bibitem{prendergast_equilibrium_1956}
K.~H. Prendergast, The {Equilibrium} of a {Self}-{Gravitating} {Incompressible}
  {Fluid} {Sphere} with a {Magnetic} {Field}. {I}., {\em \apj} {\bf 123}, p.
  498 (May 1956).

\bibitem{frieben_equilibrium_2012}
J.~Frieben and L.~Rezzolla, Equilibrium models of relativistic stars with a
  toroidal magnetic field, {\em \mnras} {\bf 427}, p. 3406 (December 2012).

\bibitem{ciolfi_2013}
R.~{Ciolfi} and L.~{Rezzolla}, {Twisted-torus configurations with large
  toroidal magnetic fields in relativistic stars.}, {\em \mnras} {\bf 435}, L43
  (Aug 2013).

\bibitem{uryu_equilibrium_2014}
K.~Uryū, E.~Gourgoulhon, C.~M. Markakis, K.~Fujisawa, A.~Tsokaros and
  Y.~Eriguchi, Equilibrium solutions of relativistic rotating stars with mixed
  poloidal and toroidal magnetic fields, {\em \prd} {\bf 90}, p. 101501
  (November 2014).

\bibitem{pili_axisymmetric_2014}
A.~G. Pili, N.~Bucciantini and L.~Del~Zanna, Axisymmetric equilibrium models
  for magnetized neutron stars in general relativity under the conformally flat
  condition, {\em \mnras} {\bf 439}, 3541  (2014).

\bibitem{samuelsson_2007}
L.~{Samuelsson} and N.~{Andersson}, {Neutron star asteroseismology. Axial crust
  oscillations in the Cowling approximation}, {\em \mnras} {\bf 374}, 256
  (January 2007).

\bibitem{sotani_2015}
H.~{Sotani}, {Torsional oscillations of neutron stars with highly tangled
  magnetic fields}, {\em \prd} {\bf 92}, p. 104024 (November 2015).

\bibitem{page_2004}
D.~{Page}, J.~M. {Lattimer}, M.~{Prakash} and A.~W. {Steiner}, {Minimal Cooling
  of Neutron Stars: A New Paradigm}, {\em \apjs} {\bf 155}, 623 (December
  2004).

\bibitem{aguilera_2008}
D.~N. {Aguilera}, J.~A. {Pons} and J.~A. {Miralles}, {The Impact of Magnetic
  Field on the Thermal Evolution of Neutron Stars}, {\em \apjl} {\bf 673}, p.
  L167 (February 2008).

\bibitem{haskell_2008}
B.~{Haskell}, L.~{Samuelsson}, K.~{Glampedakis} and N.~{Andersson}, {Modelling
  magnetically deformed neutron stars}, {\em \mnras} {\bf 385}, 531 (March
  2008).

\bibitem{gomes_2019}
R.~O. {Gomes}, H.~{Pais}, V.~{Dexheimer}, C.~{Provid{\^e}ncia} and
  S.~{Schramm}, {Limiting magnetic field for minimal deformation of a
  magnetized neutron star}, {\em \aap} {\bf 627}, p. A61 (July 2019).

\bibitem{kandel_2020}
D.~Kandel and R.~W. Romani, Atmospheric circulation on black widow companions,
  {\em \apj} {\bf 892}, p. 101 (apr 2020).

\bibitem{abbott_2018_1}
T.~L.~S. Collaboration and the Virgo~Collaboration, Gw170817: Measurements of
  neutron star radii and equation of state, {\em \prl} {\bf 121}, p. 161101
  (Oct 2018).

\bibitem{Bauswein_2019}
A.~Bauswein, Equation of state constraints from multi-messenger observations of
  neutron star mergers, {\em Ann. Physics} {\bf 411}, p. 167958 (Dec 2019).

\bibitem{abbott_gw170817:_2017}
T.~L.~S. Collaboration and V.~Collaboration, {GW}170817: {Observation} of
  {Gravitational} {Waves} from a {Binary} {Neutron} {Star} {Inspiral}, {\em
  \prl} {\bf 119}, p. 161101 (October 2017).

\bibitem{pang_2021}
P.~T.~H. {Pang}, I.~{Tews}, M.~W. {Coughlin}, M.~{Bulla}, C.~{Van Den Broeck}
  and T.~{Dietrich}, {Nuclear-Physics Multi-Messenger Astrophysics Constraints
  on the Neutron-Star Equation of State: Adding NICER's PSR J0740+6620
  Measurement}, {\em arXiv e-prints} , p. arXiv:2105.08688 (May 2021).

\bibitem{zhang_2021}
N.-B. {Zhang} and B.-A. {Li}, {Impacts of NICER's Radius Measurement of PSR
  J0740+6620 on Nuclear Symmetry Energy at Suprasaturation Densities}, {\em
  arXiv e-prints} , p. arXiv:2105.11031 (May 2021).

\bibitem{ruderman_1995}
M.~{Ruderman}, {Spin-driven changes in neutron star magnetic fields.}, {\em J.
  Astrophys. Astron.} {\bf 16}, 207 (June 1995).

\bibitem{zdunik_2013}
J.~L. {Zdunik} and P.~{Haensel}, {Maximum mass of neutron stars and strange
  neutron-star cores}, {\em \aap} {\bf 551}, p. A61 (March 2013).

\bibitem{costa_2014}
P.~{Costa}, M.~{Ferreira}, H.~{Hansen}, D.~P. {Menezes} and
  C.~{Provid{\^e}ncia}, {Phase transition and critical end point driven by an
  external magnetic field in asymmetric quark matter}, {\em \prd} {\bf 89}, p.
  056013 (March 2014).

\bibitem{cai_2015}
B.-J. {Cai}, F.~J. {Fattoyev}, B.-A. {Li} and W.~G. {Newton}, {Critical density
  and impact of {\ensuremath{\Delta}} (1232 ) resonance formation in neutron
  stars}, {\em \prc} {\bf 92}, p. 015802 (July 2015).

\bibitem{Drago_Lavagno_Pagliara_Pigato_2016}
A.~Drago, A.~Lavagno, G.~Pagliara and D.~Pigato, The scenario of two families
  of compact stars, {\em \epja} {\bf 52}, p.~40 (Feb 2016).

\bibitem{bocquet_rotating_1995}
M.~Bocquet, S.~Bonazzola, E.~Gourgoulhon and J.~Novak, Rotating neutron star
  models with a magnetic field., {\em \aap} {\bf 301}, p. 757 (September 1995).

\bibitem{cutler_2002}
C.~Cutler, Gravitational waves from neutron stars with large toroidal $b$
  fields, {\em \prd} {\bf 66}, p. 084025 (Oct 2002).

\bibitem{dallosso_2021}
S.~{Dall'Osso} and L.~{Stella}, {Millisecond Magnetars}, {\em arXiv e-prints} ,
  p. arXiv:2103.10878 (March 2021).

\bibitem{papantonopoulos_modifications_2015}
E.~Papantonopoulos, {\em Modifications of {Einstein}'s {Theory} of {Gravity} at
  {Large} {Distances}}Lecture {Notes} in {Physics}, Lecture {Notes} in
  {Physics} (Springer International Publishing, 2015).

\bibitem{trimble_existence_1987}
V.~Trimble, Existence and {Nature} of {Dark} {Matter} in the {Universe}, {\em
  \araa} {\bf 25}, 425 (September 1987).

\bibitem{peebles_cosmological_2003}
P.~J.~E. Peebles and B.~Ratra, The cosmological constant and dark energy, {\em
  \rmp} {\bf 75}, 559 (April 2003).

\bibitem{capozziello_extended_2011}
S.~Capozziello and M.~de~Laurentis, Extended {Theories} of {Gravity}, {\em
  \prep} {\bf 509}, 167  (2011).

\bibitem{damour_runaway_2002}
T.~Damour, F.~Piazza and G.~Veneziano, Runaway {Dilaton} and {Equivalence}
  {Principle} {Violations}, {\em \prl} {\bf 89}, p. 081601 (August 2002).

\bibitem{defelice_2006}
A.~{DeFelice}, M.~{Hindmarsh} and M.~{Trodden}, {Ghosts, instabilities, and
  superluminal propagation in modified gravity models}, {\em \jcap} {\bf 2006},
  p. 005 (Aug 2006).

\bibitem{defelice_2010}
A.~{De Felice} and T.~{Tanaka}, {Inevitable Ghost and the Degrees of Freedom in
  f(R,G) Gravity}, {\em \ptp} {\bf 124}, 503 (Sep 2010).

\bibitem{bertolami_2016}
O.~{Bertolami} and J.~{P{\'a}ramos}, {Viability of nonminimally coupled f (R)
  gravity}, {\em \grg} {\bf 48}, p.~34 (Mar 2016).

\bibitem{damour_nonperturbative_1993}
T.~Damour and G.~Esposito-Farèse, Nonperturbative strong-field effects in
  tensor-scalar theories of gravitation, {\em \prl} {\bf 70}, 2220 (April
  1993).

\bibitem{eardley_1973}
D.~M. Eardley, D.~L. Lee and A.~P. Lightman, Gravitational-wave observations as
  a tool for testing relativistic gravity, {\em Phys. Rev. D} {\bf 8}, 3308
  (Nov 1973).

\bibitem{pang_2020}
P.~T.~H. Pang, R.~K.~L. Lo, I.~C.~F. Wong, T.~G.~F. Li and C.~Van Den~Broeck,
  Generic searches for alternative gravitational wave polarizations with
  networks of interferometric detectors, {\em \prd} {\bf 101}, p. 104055 (May
  2020).

\bibitem{shibata_coalescence_2014}
M.~Shibata, K.~Taniguchi, H.~Okawa and A.~Buonanno, Coalescence of binary
  neutron stars in a scalar-tensor theory of gravity, {\em \prd} {\bf 89}, p.
  084005 (April 2014).

\bibitem{sotani_2005}
H.~{Sotani} and K.~D. {Kokkotas}, {Stellar oscillations in scalar-tensor theory
  of gravity}, {\em \prd} {\bf 71}, p. 124038 (June 2005).

\bibitem{pani_2014}
P.~Pani and E.~Berti, Slowly rotating neutron stars in scalar-tensor theories,
  {\em \prd} {\bf 90}, p. 024025 (Jul 2014).

\bibitem{doneva_rapidly_2014}
D.~D. Doneva, S.~S. Yazadjiev, N.~Stergioulas and K.~D. Kokkotas, Rapidly
  rotating neutron stars in scalar-tensor theories of gravity, {\em \prd} {\bf
  88}, p. 084060 (Oct 2013).

\bibitem{soldateschi_2020}
J.~{Soldateschi}, N.~{Bucciantini} and L.~{Del Zanna}, {Axisymmetric
  equilibrium models for magnetised neutron stars in scalar-tensor theories},
  {\em \aap} {\bf 640}, p. A44 (August 2020).

\bibitem{soldateschi_2021}
J.~{Soldateschi}, N.~{Bucciantini} and L.~{Del Zanna}, Magnetic deformation of
  neutron stars in scalar-tensor theories, {\em A\&A} {\bf 645}, p. A39
  (2021).

\bibitem{soldateschi_2021_1}
J.~{Soldateschi}, N.~{Bucciantini} and L.~{Del Zanna}, {Quasi-universality of
  the magnetic deformation of neutron stars in general relativity and beyond},
  {\em arXiv e-prints} , p. arXiv:2106.00603 (June 2021).

\bibitem{bucciantini_2020}
N.~{Bucciantini} and J.~{Soldateschi}, {Iron line from neutron star accretion
  discs in scalar tensor theories}, {\em \mnras} {\bf 495}, L56 (April 2020).

\bibitem{will_confrontation_2014}
C.~M. Will, The {Confrontation} between {General} {Relativity} and
  {Experiment}, {\em \lrr} {\bf 17}  (2014).

\bibitem{shao_constraining_2017}
L.~Shao, N.~Sennett, A.~Buonanno, M.~Kramer and N.~Wex, Constraining
  nonperturbative strong-field effects in scalar-tensor gravity by combining
  pulsar timing and laser-interferometer gravitational-wave detectors, {\em
  \prx} {\bf 7}, 041025 (October 2017).

\bibitem{voisin_2020}
G.~{Voisin}, I.~{Cognard}, P.~C.~C. {Freire}, N.~{Wex}, L.~{Guillemot},
  G.~{Desvignes}, M.~{Kramer} and G.~{Theureau}, {An improved test of the
  strong equivalence principle with the pulsar in a triple star system}, {\em
  \aap} {\bf 638}, p. A24 (June 2020).

\bibitem{ramazanoglu_spontaneous_2016}
F.~M. Ramazanoğlu and F.~Pretorius, Spontaneous {Scalarization} with {Massive}
  {Fields}, {\em \prd} {\bf 93}, p. 064005 (March 2016).

\bibitem{yaza_2016}
S.~S. {Yazadjiev}, D.~D. {Doneva} and D.~{Popchev}, {Slowly rotating neutron
  stars in scalar-tensor theories with a massive scalar field}, {\em \prd} {\bf
  93}, p. 084038 (Apr 2016).

\bibitem{rosca_2020}
R.~Rosca-Mead, C.~J. Moore, U.~Sperhake, M.~Agathos and D.~Gerosa, {Structure
  of neutron stars in massive scalar-tensor gravity}, {\em Symmetry} {\bf 12},
  p. 1384  (2020).

\bibitem{oron_relativistic_2002}
A.~Oron, Relativistic magnetized star with poloidal and toroidal fields, {\em
  \prd} {\bf 66}, p. 023006 (July 2002).

\bibitem{pili_general_2017}
A.~G. Pili, N.~Bucciantini and L.~Del~Zanna, General relativistic models for
  rotating magnetized neutron stars in conformally flat space-time, {\em
  \mnras} {\bf 470}, 2469  (2017).

\bibitem{wilson_mathews_2003}
J.~R. Wilson and G.~J. Mathews, {\em Relativistic Numerical
  Hydrodynamics}Cambridge Monographs on Mathematical Physics, Cambridge
  Monographs on Mathematical Physics (Cambridge University Press, 2003).

\bibitem{isenberg_waveless_2008}
J.~A. Isenberg, Waveless approximation theories of gravity, {\em \ijmpd} {\bf
  17}, 265 (February 2008).

\bibitem{bucciantini_fully_2013}
N.~Bucciantini and L.~Del~Zanna, A fully covariant mean-field dynamo closure
  for numerical 3 + 1 resistive {GRMHD}, {\em \mnras} {\bf 428}, 71 (January
  2013).

\bibitem{tomei2020}
N.~{Tomei}, L.~{Del Zanna}, M.~{Bugli} and N.~{Bucciantini}, {General
  relativistic magnetohydrodynamic dynamo in thick accretion discs: fully
  non-linear simulations}, {\em \mnras} {\bf 491}, 2346 (Jan 2020).

\bibitem{alcubierre_introduction_2008}
M.~Alcubierre, {\em Introduction to 3+1 Numerical Relativity}International
  Series of Monographs on Physics, International Series of Monographs on
  Physics (OUP Oxford, 2008).

\bibitem{gourgoulhon_3+1_2012}
{\'E}.~Gourgoulhon, {\em 3+1 Formalism in General Relativity: Bases of
  Numerical Relativity}Lecture Notes in Physics, Lecture Notes in Physics
  (Springer Berlin Heidelberg, 2012).

\bibitem{del_zanna_exact_1996}
L.~{Del Zanna} and C.~{Chiuderi}, {Exact solutions for symmetric
  magnetohydrodynamic equilibria with mass flow.}, {\em \aap} {\bf 310}, 341
  (June 1996).

\bibitem{bergmann_comments_1968}
P.~G. Bergmann, Comments on the scalar-tensor theory, {\em \ijtp} {\bf 1}, 25
  (May 1968).

\bibitem{wagoner_scalar-tensor_1970}
R.~V. Wagoner, Scalar-{Tensor} {Theory} and {Gravitational} {Waves}, {\em \prd}
  {\bf 1}, 3209 (June 1970).

\bibitem{santiago_2000}
D.~I. Santiago and A.~S. Silbergleit, On the energy-momentum tensor of the
  scalar field in scalar-tensor theories of gravity, {\em \grg} {\bf 32}, p.
  565–582 (Apr 2000).

\bibitem{doneva_rapidly_2016}
D.~D. Doneva and S.~S. Yazadjiev, Rapidly rotating neutron stars with a massive
  scalar field{\textemdash}structure and universal relations, {\em \jcap} {\bf
  2016}, 019 (nov 2016).

\bibitem{zhang_gravitational_2017}
X.~Zhang, T.~Liu and W.~Zhao, Gravitational radiation from compact binary
  systems in screened modified gravity, {\em \prd} {\bf 95}, p. 104027 (May
  2017), arXiv: 1702.08752.

\bibitem{zhang_2019}
X.~Zhang, R.~Niu and W.~Zhao, Constraining the scalar-tensor gravity theories
  with and without screening mechanisms by combined observations, {\em \prd}
  {\bf 100}, p. 024038 (Jul 2019).

\bibitem{pili_general_2015}
A.~G. Pili, N.~Bucciantini and L.~Del~Zanna, General relativistic neutron stars
  with twisted magnetosphere, {\em \mnras} {\bf 447}, 2821 (March 2015).

\bibitem{bucciantini_role_2015}
N.~Bucciantini, A.~G. Pili and L.~D. Zanna, The role of currents distribution
  in general relativistic equilibria of magnetized neutron stars, {\em \mnras}
  {\bf 447}, 1  (2015).

\bibitem{bucciantini_general_2011}
N.~Bucciantini and L.~Del~Zanna, General relativistic magnetohydrodynamics in
  axisymmetric dynamical spacetimes: the {X}-{ECHO} code, {\em \aap} {\bf 528},
  A101  (2011).

\bibitem{wentzel_1960}
D.~G. {Wentzel}, {Hydromagnetic Equilibria.}, {\em \apjs} {\bf 5}, p. 187
  (December 1960).

\bibitem{ostriker_1969}
J.~P. {Ostriker} and J.~E. {Gunn}, {On the Nature of Pulsars. I. Theory}, {\em
  \apj} {\bf 157}, p. 1395 (September 1969).

\bibitem{Fonseca_2021}
E.~{Fonseca}, H.~T. {Cromartie}, T.~T. {Pennucci}, P.~S. {Ray}, A.~Y.
  {Kirichenko}, S.~M. {Ransom}, P.~B. {Demorest}, I.~H. {Stairs},
  Z.~{Arzoumanian}, L.~{Guillemot}, A.~{Parthasarathy}, M.~{Kerr},
  I.~{Cognard}, P.~T. {Baker}, H.~{Blumer}, P.~R. {Brook}, M.~{DeCesar},
  T.~{Dolch}, F.~A. {Dong}, E.~C. {Ferrara}, W.~{Fiore}, N.~{Garver-Daniels},
  D.~C. {Good}, R.~{Jennings}, M.~L. {Jones}, V.~M. {Kaspi}, M.~T. {Lam}, D.~R.
  {Lorimer}, J.~{Luo}, A.~{McEwen}, J.~W. {McKee}, M.~A. {McLaughlin},
  N.~{McMann}, B.~W. {Meyers}, A.~{Naidu}, C.~{Ng}, D.~J. {Nice}, N.~{Pol},
  H.~A. {Radovan}, B.~{Shapiro-Albert}, C.~M. {Tan}, S.~P. {Tendulkar}, J.~K.
  {Swiggum}, H.~M. {Wahl} and W.~W. {Zhu}, {Refined Mass and Geometric
  Measurements of the High-mass PSR J0740+6620}, {\em \apjl} {\bf 915}, p. L12
  (July 2021).

\bibitem{Fortin_Providencia_Raduta_Gulminelli_Zdunik_Haensel_Bejger_2016}
M.~Fortin, C.~Providencia, A.~R. Raduta, F.~Gulminelli, J.~L. Zdunik,
  P.~Haensel and M.~Bejger, Neutron star radii and crusts: uncertainties and
  unified equations of state, {\em Phys.Rev.C} {\bf 94}  (2016).

\bibitem{chaves_2019}
A.~{Guerra Chaves} and T.~{Hinderer}, {Probing the equation of state of neutron
  star matter with gravitational waves from binary inspirals in light of
  GW170817: a brief review}, {\em J. Phys. G: Nucl. Part. Phys.} {\bf 46}, p.
  123002 (December 2019).

\bibitem{bauswein_2017}
A.~Bauswein, O.~Just, H.-T. Janka and N.~Stergioulas, Neutron-star radius
  constraints from gw170817 and future detections, {\em \apjl} {\bf 850}, p.
  L34 (Dec 2017).

\bibitem{riley_2021}
T.~E. {Riley}, A.~L. {Watts}, P.~S. {Ray}, S.~{Bogdanov}, S.~{Guillot}, S.~M.
  {Morsink}, A.~V. {Bilous}, Z.~{Arzoumanian}, D.~{Choudhury}, J.~S. {Deneva},
  K.~C. {Gendreau}, A.~K. {Harding}, W.~C.~G. {Ho}, J.~M. {Lattimer},
  M.~{Loewenstein}, R.~M. {Ludlam}, C.~B. {Markwardt}, T.~{Okajima},
  C.~{Prescod-Weinstein}, R.~A. {Remillard}, M.~T. {Wolff}, E.~{Fonseca}, H.~T.
  {Cromartie}, M.~{Kerr}, T.~T. {Pennucci}, A.~{Parthasarathy}, S.~{Ransom},
  I.~{Stairs}, L.~{Guillemot} and I.~{Cognard}, {A NICER View of the Massive
  Pulsar PSR J0740+6620 Informed by Radio Timing and XMM-Newton Spectroscopy},
  {\em arXiv e-prints} , p. arXiv:2105.06980 (May 2021).

\bibitem{breu_2016}
C.~{Breu} and L.~{Rezzolla}, {Maximum mass, moment of inertia and compactness
  of relativistic stars}, {\em \mnras} {\bf 459}, 646 (June 2016).

\bibitem{atnf_2005}
R.~N. {Manchester}, G.~B. {Hobbs}, A.~{Teoh} and M.~{Hobbs}, {The Australia
  Telescope National Facility Pulsar Catalogue}, {\em \aj} {\bf 129}, 1993
  (April 2005).

\bibitem{antoniadis_2016}
J.~{Antoniadis}, T.~M. {Tauris}, F.~{Ozel}, E.~{Barr}, D.~J. {Champion} and
  P.~C.~C. {Freire}, {The millisecond pulsar mass distribution: Evidence for
  bimodality and constraints on the maximum neutron star mass}, {\em arXiv
  e-prints} , p. arXiv:1605.01665 (May 2016).

\bibitem{faucher_2006}
C.-A. {Faucher-Gigu{\`e}re} and V.~M. {Kaspi}, {Birth and Evolution of Isolated
  Radio Pulsars}, {\em \apj} {\bf 643}, 332 (May 2006).

\end{thebibliography}

\end{document}